\documentclass{AIAA}
\usepackage{amsmath,amsthm,amssymb,graphicx,bm,url}

\newcommand{\obs}{\text{obs}}

\newcommand{\MtoT}{\text{M-to-T}}
\newcommand{\TtoM}{\text{T-to-M}}

\newcommand{\MtoE}{\text{M-to-E}}

\newcommand{\nil}[1]{}

\begin{document}

\title{Observation-centered Kalman filters}

\author{John T. Kent\footnote{Professor, Department of Statistics, j.t.kent@leeds.ac.uk.} and Shambo Bhattacharjee\footnote{PhD student, Department of Statistics, mmsb@leeds.ac.uk.}}
\affiliation{University of Leeds, Leeds, UK}
\author{Weston R. Faber\footnote{Research Scientist, weston.faber@l3harris.com.}}
\affiliation{L3Harris, Applied Defense Solutions, Colorado Springs, CO, USA}
\author{Islam I. Hussein\footnote{Senior R\&D Scientist, islam.Hussein@l3harris.com.}}
\affiliation{L3Harris, Applied Defense Solutions, Herndon, VA, USA}

\begin{abstract}
  Various methods have been proposed for the nonlinear filtering
  problem, including the extended Kalman filter (EKF), iterated
  extended Kalman filter (IEKF), unscented Kalman filter (UKF) and
  iterated unscented Kalman filter (IUKF).  In this paper two new
  nonlinear Kalman filters are proposed and investigated, namely the
  observation-centered extended Kalman filter (OCEKF) and
  observation-centered unscented Kalman filter (OCUKF). Although the
  UKF and EKF are common default choices for nonlinear filtering,
  there are situations where they are bad choices.  Examples are given
  where the EKF and UKF perform very poorly, and the IEKF and OCEKF
  perform well.  In addition the IUKF and OCUKF are generally similar
  to the IEKF and OCEKF, and also perform well, though care is needed
  in the choice of tuning parameters when the observation error is
  small. The reasons for this behaviour are explored in detail.
\end{abstract}

\maketitle

\section{Introduction}
\label{sec:intro}
Consider a system of state and measurement (or observation) equations
where (a) the propagation equation is linear, but (b) the measurement
equation is nonlinear, and (c) the measurement or observation variance
is small.  Due to the nonlinearity, the classic Kalman filter (KF) is
not defined, so various modifications have been proposed, including
the extended Kalman filter (EKF), the iterated extended Kalman filter
(IEKF), the unscented Kalman filter (UKF), and the iterated unscented
Kalman filter (IUKF).  To this collection two new filters are added
here and given the names \emph{observation-centered extended Kalman
filter (OCEKF)} and \emph{observation-centered unscented Kalman filter
(OCUKF)}.

In this paper these filters are explored in a one-dimensional
setting. In particular, it is shown that in certain circumstances with
small measurement error,
\begin{itemize}
\item the EKF and UKF perform very poorly;

\item the IEKF and OCEKF are similar to one another and
  perform well; 

\item the IUKF and OCUKF is generally similar to the IEKF and OCEKF,
  and also perform well, but if the measurement error is very small,
  care is needed in the choice of tuning parameters.

\end{itemize}

Although this paper is set in the context of a one-dimensional state
and a one-dimensional observation, the ideas are also relevant in
higher dimensions.  An important example is given by the tracking
problem in orbital dynamics.  The state of an orbiting object is
six-dimensional and an angles-only observation is two-dimensional.
However, the main source of uncertainty in the state is often
one-dimensional, given by the position of the object along the orbital
path.  The other five state variables do not change with time under
Keplerian dynamics.

It is convenient to describe the position of the object along the
orbital path using the \emph{mean anomaly} because this choice makes
the propagation equation linear.  However, the projection of an
angles-only observation onto the orbital plane is measured in terms of
\emph{true anomaly}.  For a highly eccentric orbit the mapping between
these two variables can be very nonlinear.  A one-dimensional version
of the tracking problem is studied in detail in Section \ref{example}.

\section{Background}
\label{sec:back}
\subsection{The classic Kalman filter} 
\label{sec:back:classic} 
Recall the classic Kalman filter~\cite{kf}, which is designed for
linear propagation and observation equations, with Gaussian noise.
There is a sequence of $p$-dimensional state vectors $\{\bm x_k\}$ and
a sequence of $q$-dimensional noisy (partial) observations
$\{\bm z_k\}$ at times $t_k, \ k \geq 1$.  Let $\mathcal{F}_k$ denote
the information contained in the first $k$ observations
$\bm z_1 , \ldots, \bm z_k$.  The state vectors evolve through noisy
linear propagation
\begin{equation}
\label{eq:classic-system}
\bm x_k = F_k \bm x_{k-1} + \bm w_k,
\end{equation}
where $F_k$ is a $p \times p$ matrix and $\bm w_k$ is system noise.  The observations are noisy versions
of linear functions of the state vectors,
\begin{equation}
\label{eq:classic-observation}
\bm z_k = H_k \bm x_k + \bm v_k,
\end{equation}
where $H_k$ is a $q \times p$ matrix and $\bm v_k$ is measurement
noise.  The random vectors $\bm w_k$ and $\bm v_k$ are assumed
independent of one another and of $\bm z_1 , \ldots, \bm z_{k-1}$,
with $N_p(0, Q_k)$ and $N_q(0,R_k)$ distributions, respectively.  The
dimension $p$ of the state vector is allowed to be different from the
dimension $q$ of the observation vector.

Start with an initial Gaussian distribution for $\bm x_0$, with mean
vector and covariance matrix denoted $\bm x_{0|0}, P_{0|0}$.  Then the
conditional propagated distribution of $\bm x_{k}$ given $\mathcal{F}_{k}$,
$k \ge 1$ follows a Gaussian distribution.  The conditional mean
vector and covariance matrix, denoted $\bm x_{k|k}, P_{k|k}$, say, can
be determined iteratively as follows.
 
Suppose $\bm x_{k-1|k-1}$ and $P_{k-1|k-1}$ are known.  After
propagation from time $t_{k-1}$ to $t_k$, the conditional distribution
of the state becomes
$$
\bm x_{k}|\mathcal{F}_{k-1} \sim N_p(\bm x_{k|k-1}, P_{k|k-1}),
$$
where
$$
\bm x_{k|k-1} = F_k \bm x_{k-1|k-1}, \quad P_{k|k-1}= F_k P_{k-1|k-1} F_k^T + Q_k.
$$

Given the observation $\bm z_k$ at time $t_k$ the Bayesian update yields the
posterior distribution
$$
\bm x_{k}|\mathcal{F}_{k} \sim N(\bm x_{k|k}, P_{k|k})
$$
with updated mean vector and covariance matrix 
\begin{align}
x_{k|k} &= \left(P_{k|k-1}^{-1} +H^T_k R_k^{-1}H_k\right)^{-1}
\left(P_{k|k-1}^{-1} \bm x_{k|k-1} +H_k^T R_k^{-1} \bm z_k \right), 
\label{eq:kf-mean}\\
P_{k|k} &= \left(P_{k|k-1}^{-1} +H^T_k R_k^{-1}H_k\right)^{-1}.
\label{eq:kf-var}
\end{align}
These expressions can also be written as 
\begin{align*}
x_{k|k} & = x_{k|k-1} + K_k (z_k - H_k x_{k|k-1}) \\
P_{k|k} &= (I - K_k H_k)P_{k|k-1}(I - K_k H_k)^T + K_kR_kK_k^T,
\end{align*}
where
\begin{align*}
K_k = P_{k|k-1}H_k^T (H_kP_{k|k-1}H_k^T +R_k)^{-1}
\end{align*}

\subsection{The EKF, IEKF and OCEKF}
\label{sec:back:ekfs}

The EKF was created to accommodate certain sorts of nonlinearity to
generalize either $F_k$ in the propagation equation or $H_k$ in the
observation equation. This paper is focused on the second situation in
a very particular setting, namely
\begin{itemize}
\item the state and observation are one-dimensional, $p=q=1$,
\item the propagation equation is exactly linear as in
  (\ref{eq:classic-system}), but
\item the measurement equation is not linear so a generalization of
  (\ref{eq:classic-observation}) is needed.
\end{itemize}

The essential ingredients for one update step can be written in a
more concise notation as follows:
\begin{align}
x \sim N(\mu_x, \sigma^2) \label{eq:update-prior}\\
z|x \sim N(h(x), \tau^2) \label{eq:update-likelihood},
\end{align}
where $h(\cdot)$ is a known monotone function, and where
$x, \ \mu_x, \ \sigma^2, \ \tau^2$ correspond to $x_k$, $x_{k|k-1}$,
$P_{k|k-1}$, $R_k$, respectively, in the previous section.  The
explicit conditioning on $\mathcal{F}_{k-1}$ in the notation is
dropped since everything is conditioned on it.  Further, boldface is
dropped since the vectors are scalars here.  Also, it is possible to
move back and forth between the state space and the observation space
by writing $z(x) = h(x)$ for any $x$ and $x(z) = h^{-1}(z)$ for any
$z$.

Equation (\ref{eq:update-prior}) can be regarded as a ``prior'' distribution
for $x$, and (\ref{eq:update-likelihood}) as the ``likelihood'' for the
observation $z$ given $x$.  Hence the posterior distribution for $x$
given a realization $z_\obs$ of the observation $z$ is proportional to
\begin{equation}
\label{eq:post}
f(x|z_\obs) \propto \exp\left\{ -\frac12 \frac{(x-\mu_x)^2}{\sigma^2}
- \frac12 \frac{(z_\obs-h(x))^2}{\tau^2} \right\}.
\end{equation}
The purpose of the update step in the EKF is to approximate this
posterior distribution by a Gaussian distribution.   Write
\begin{align}
z_\obs-h(x) &= z_\obs-h(y) + h(y)-h(x) \notag\\
&\approx z_\obs-h(y) + h'(y)(y-x) \label{eq:taylor},
\end{align}
using a first order Taylor series expansion, where the choice of $y$
is discussed below.  Then the exponent in (\ref{eq:post}) becomes a
quadratic function of $x$; hence the approximating posterior
distribution is Gaussian with mean and variance
\begin{align}
\mu_{x|z_\obs} &= \mu_x + \frac{h' \sigma^2}{h'^2 \sigma^2+ \tau^2} 
\{z_\obs - h(y) + h'[y - \mu_x] \}, 
\label{eq:ekfs-mean}\\
\sigma^2_{x|z_\obs} &=  \left(\frac{1}{\sigma^2} + 
\frac{h'^2}{\tau^2}\right)^{-1}, \label{eq:ekf-var}
\end{align}
where $h' = h'(y)$.

There are three important choices for $y$.
\begin{itemize}
\item[(a)] $y=\mu_x$, the prior mean.  This choice gives the standard EKF
 ~\cite{Havl_k_2015, wiki}.  Equation (\ref{eq:ekfs-mean}) for the
  posterior mean simplifies to
\begin{equation}
\mu_{x|z_\obs} = \mu_x + \frac{h' \sigma^2}{h'^2 \sigma^2+ \tau^2} 
\left\{z_\obs - h(\mu_x) \right\} \label{eq:ekf-mean}.
\end{equation}

\item[(b)] $y=\mu_{x|z_\obs}$, the posterior mean.  This choice gives the
  iterated EKF~\cite{Havl_k_2015}.  Equation (\ref{eq:ekfs-mean}) for
  the posterior mean becomes
\begin{equation}
\mu_{x|z_\obs} = \mu_x + \frac{h' \sigma^2}{h'^2 \sigma^2+ \tau^2} 
\left\{z_\obs - h(\mu_{x|z_\obs}) + h'[\mu_{x|z_\obs} - \mu_x] \right\}.
\label{eq:iekf-mean}
\end{equation}
Note that $\mu_{x|z_\obs}$ occurs on both sides of the equation.
Hence an iterative algorithm is needed to compute it.

\item[(c)] $y=x_\obs = h^{-1}(z_\obs)$, the transformed
  observation. This choice gives the new proposal of this paper, the
observation-centered EKF.  Equation
  (\ref{eq:ekfs-mean}) for the posterior mean simplifies to
\begin{equation}
\mu_{x|z_\obs} = \mu_x + \frac{h'^2 \sigma^2}{h'^2 \sigma^2+ \tau^2} 
\left\{x_\obs - \mu_x \right\} \label{eq:ocekf-mean}.
\end{equation}
\end{itemize}

\subsection{The UKF, IUKF and OCUKF}
\label{sec:back:ukfs}
Various unscented filters (UKF, IUKF and OCUKF) can be defined  as
discrete approximations to the extended filters (EKF, IEKF and OCEKF).
The starting point is a collection of three ``sigma points''
\begin{equation}
\label{eq:sigma-points}
x_{-1}=y-\alpha \sigma, \quad x_0=y, \quad x_1=y+\alpha \sigma,
\end{equation}
where $y$ is a centering point to be specified, and $\alpha>0$ is a tuning
parameter.  Two sets of
weights are defined,
\begin{align*}
w^a_{-1} &= w^a_1 = \frac{1}{2 \alpha^2}, \quad w^a_0 = 1-\frac{1}{\alpha^2}\\
w^c_{-1}&=w^c_1 = \frac{1}{2 \alpha^2}, \quad w^c_0 = 1-\frac{1}{\alpha^2}
+3-\alpha^2.
\end{align*}
where the first weights are used to compute means and  the second weights are used
to compute variances and covariances.

The conventional choice of tuning parameter $\alpha = 0.001$ is used
here so that the sigma points are close to $y$.  The sigma points have
weighted mean and variance.
$$
\sum  w^a_j x_j = y, \quad \sum w^c_j (x_j - y)^2 = \sigma^2,
$$
where in all cases the sums range over $j=-1,0,1$.

Let $z_j = h(x_j)$ denote the transformed sigma points,
with mean $\bar{z} = \sum w^a_j z_j$.  Let $C=C(y)$ and $V=V(y)$ denote the
weighted covariance between the $\{z_j\}$ and $\{x_j\}$, and the
weighted variance of the $\{z_j\}$, respectively,
$$
C(y) = \sum w^c_j(z_j - \bar{z})(x_j - y), \quad
V(y)= \sum w^c_j (z_j - \bar{z})^2.
$$
The limiting behavior of $C,V$ and $\bar{z}$ as $\alpha \rightarrow 0$
~\cite{all-scale} is given by
\begin{equation}
\label{eq:limit}
V \rightarrow  h'(y)^2 \sigma^2, \quad
C \rightarrow  h'(y) \sigma^2, \quad
\bar{z} \rightarrow  h(y) + \frac12 \sigma^2 h''(y).
\end{equation}

Hence several choices of unscented filter can be defined by mimicking
the extended filters in (\ref{eq:ekfs-mean})--(\ref{eq:ekf-var}),
\begin{equation}
\label{eq:ukfs}
\mu_{x|z_\obs} = \mu_x + \frac{C}{V+\tau^2} \{z_\obs- h(y) + 
(V/C) [y-\mu_x]\}, \quad
\sigma^2_{x_|z_\obs} = \sigma^2 - C^2/(V + \tau^2),
\end{equation}
for suitable values of $y$.

\begin{itemize}
\item[(a)] The standard UKF~\cite{ukf_Julier, ukf_Wan} uses $y=\mu_x$.  In addition
it  
replaces  $h(y)$ by $\bar{z}$ in  (\ref{eq:ukfs}) to provide some bias correction.
 
\item [(b)] The IUKF~\cite{IUKF_Sibley} uses $y = \mu_{x|z_\obs}$ in
  (\ref{eq:ukfs}). As conventionally formulated, the IUKF does not
incorporate any bias correction.
\item [(c)] Similarly to the IUKF, it is possible to define an
  observation centered UKF (OCUKF) by using
  $y=x_\obs = h^{-1}(z_\obs)$ in (\ref{eq:ukfs}).
\end{itemize}

\section{Intuition behind the iterated and observation-centered filters}
\label{sec:intuition}
For an $N(\mu, \sigma^2)$ distribution, use the phrase \emph{effective
  range} (or more precisely 95\% effective range) to describe the
interval $(\mu - 2 \sigma, \mu+2 \sigma)$ covering 95\% of the
probability mass.

The Taylor series approximation in (\ref{eq:taylor}) will be a good
approximation if $h$ is approximately linear over an interval
containing $y$ and $x$.  When $\tau^2$ is small and $\sigma^2$ is not
small, then the posterior distribution of $x$ will be concentrated
near $x_\obs$.  The choice $y=\mu_x$ may not be a good choice in this
setting; the effective support of the posterior distribution of $x$
may be a long way from $\mu_x$ and $h$ may be very nonlinear over this
interval.  On the other hand, both $y=\mu_{x|z_\obs}$ and $y=x_\obs$
may be very good choices; these two values will be close together and
the posterior distribution will be concentrated near both these
choices.

More generally in the one-dimensional non-linear filtering system,
information can be represented on two scales: the \emph{signal scale}
$x$ and the \emph{measurement or observation scale} $z$.  The state variance
$\sigma^2$ is on the signal scale and the observation variance
$\tau^2$ is on the measurement scale.  Variability can be mapped back and
forth between the two scales by assuming $h$ is approximately linear
over an appropriate range and multiplying or dividing by $h'(\xi)$
where $\xi$ is an appropriate value within the interval (on the signal
scale) on which the signal varies.  

Hence there are two choices for linearization: the first choice is to
map the prior variability in $x$ from the signal scale to the
measurement scale, and then to combine the prior variability with the
likelihood on the measurement scale.  This is the approach effectively
taken in the standard EKF and UKF, with the posterior density finally
mapped back to the signal scale.  However, although the prior
distribution of $x$ is exactly Gaussian on the signal scale, it may be
very non-Gaussian when transformed to the measurement scale when
$\sigma^2$ is not small.

The second choice is to map the variability in $z$ about $h(x)$ from
the measurement scale to the signal scale.  Because $\tau^2$ is assumed small,
this is a reasonable approach; it is the approach taken by the iterated and observation-centered filters.

There are several features in this setup that make the iterated and
observation-centered filters feasible and effective.
\begin{itemize}
\item The prior distribution of the signal is exactly
normal.
\item The transformation function $h$ is allowed to be highly nonlinear.
\item But the standard deviation $\tau$ for the distribution of an
observation $z$ given $x$ is small.
\item The mapping between signal space and observation space is
  one to one.  In particular it is possible to define $x_\obs$, the
  value on the signal scale corresponding to the observation $z_\obs$
  on the measurement scale.
\end{itemize}

\section{Idealized analytic examples}
\label{sec:ideal}
To illustrate the issues involved, consider an idealized version of
the problem and limit attention to the extended filters.  Suppose
$h(x) = x^\lambda$ is the mapping from the signal scale to the
measurement scale, where $\lambda$ is a known power.  Let $\mu_x = 1$
and let $\tau^2=0$, so there is no measurement error.  The choice of
$\sigma^2$ is irrelevant for this section.  Let $z_\obs = 2$ be the
realized value of the observation.  Since $\tau^2 = 0$ the correct
posterior distribution is concentrated at
$\mu_{x|z}= h^{-1}(2) = 2^{1/\lambda}$.

The standard EKF gets the posterior variance $\sigma^2_{x|z_\obs} = 0$
correct, but gets the posterior mean wrong. Here are the details.  For
the standard EKF,
$$
\mu_{x|z_\obs}=1 + 1/\lambda.
$$
Results for standard EKF and IEKFs are summarized in Table 1.  The table also
includes the results for the observation-centered EKF, defined below.
Note that the EKF overshoots the exact posterior mean if $\lambda>1$ and
undershoots the exact posterior mean if $\lambda<1$. The IEKF and OCEKF results match the exact result.

\begin{table}[h!]
\begin{center}
\caption{Comparison between various approximations to the posterior mean for the idealized example in Section \ref{sec:ideal}. In each case the posterior variance is 0.} 
\vspace{2 mm}
\label{my-label11}
\begin{tabular}{ccc} 
\hline
         &     & IEKF  \\ 
$\lambda$    & EKF & OCEKF  \\
         &     & Exact  \\ \hline
1      & 2        & 2.00  \\ 
2      & 1.5     & 1.41   \\ 
0.5    & 3       & 4     \\ \hline
\end{tabular} \end{center} 
\end{table}

\section{Application to 1d orbital dynamics} \label{example}
\label{sec:orbital}
Consider a small body in orbit about a large body, such as a satellite
about the earth.  Suppose the object follows Keplerian dynamics, so
that it follows an exact elliptical orbit.  For simplicity suppose the
orbital plane, the angle of perigee and the ellipticity are known
exactly.  Then the orbit can be represented in the $x-y$ plane, with
the angle of perigee at 0$^o$, pointing to the positive $x$-axis.

There are three angles of mathematical interest in this setting to
describe the position of the object along the orbit: the
\emph{eccentric anomaly} ($E$), the \emph{mean anomaly} ($M$) and the
\emph{true anomaly} ($T$), where all three angles are measured from
perigee. The true anomaly describes the actual angular position of the
object, as measured from the center of the earth.  The mean anomaly
has constant derivative with respect to time as the object moves, and
hence simplifies the mathematical development.  The eccentric anomaly
is an intermediate angle of no direct interest.  The relation between
the angles is given as follows in radians, where $e$ is the
ellipticity, $0 \leq e < 1$:
\begin{align*}
\tan \frac12 T &= \sqrt{\frac{1+e}{1-e}} \tan \frac12 E,\\
M &= E - e \sin E.
\end{align*}
These mappings are bijective, so any one angle determines the other
two.  The calculations are all straightforward, except that a
numerical iteration is needed to solve for $E$ from $M$.

Initially all three angles are defined on the same interval
$-\pi \leq E,M,T \leq \pi$.  The angles agree at the midpoint and
endpoints.  That is, if $E=0,\pi$ or $-\pi$, then $M$ and $T$ also
equal to $0,\pi$ or $-\pi$, respectively.  Further the identification
between angles is symmetric about the origin.  That is, if $E$
corresponds to $M$ and $T$, then $-E$ corresponds to $-M$ and $-T$.
Finally, by periodic extension, the mapping between the three angles
can be extended to any interval
$-\pi + 2\pi k \leq E,M,T \leq \pi +2 \pi k$, $k \in \mathbb{Z}$,
and thus to the whole real line.

\begin{figure}[ht]
\includegraphics[width=8.2cm, keepaspectratio]{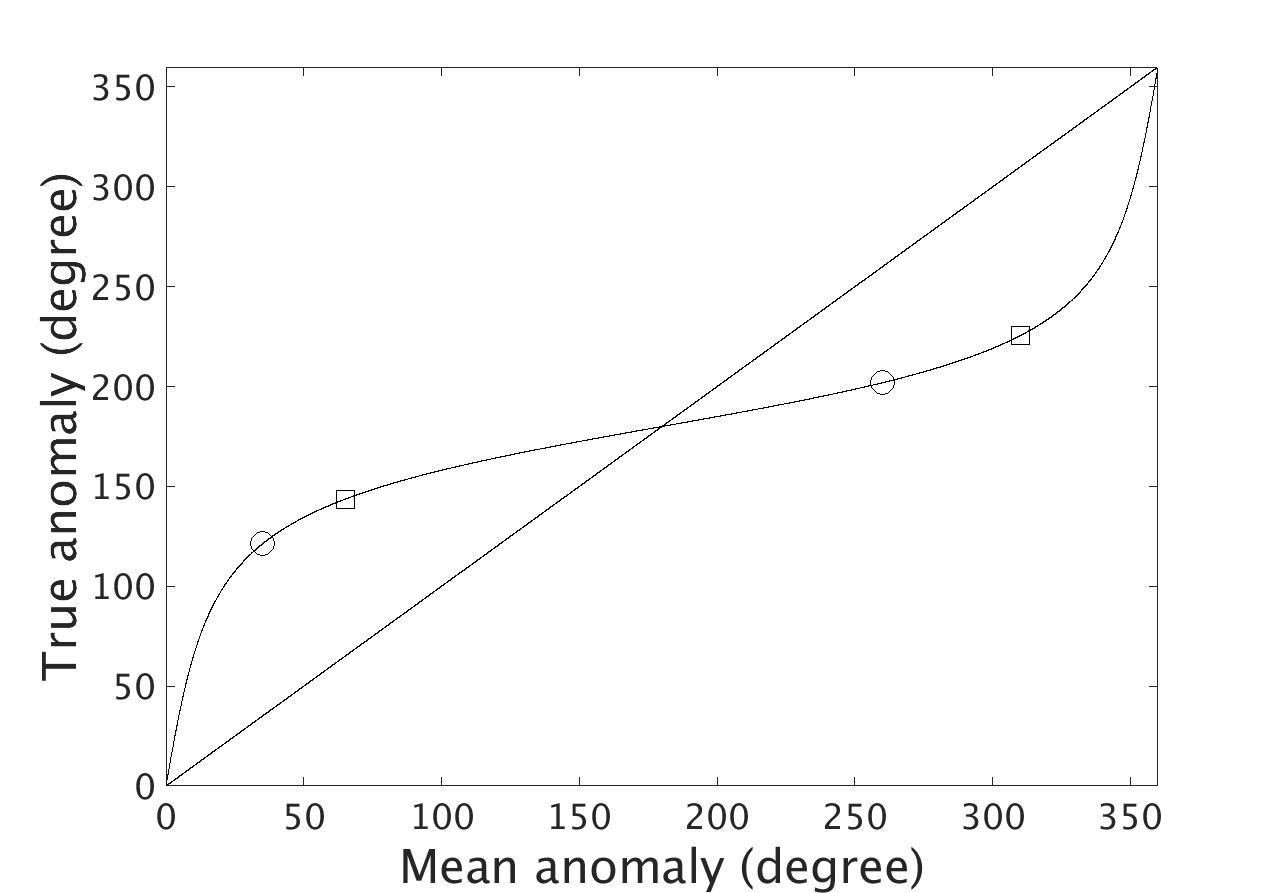}
\caption{True anomaly as a function of mean anomaly, for eccentricity
  $e=0$ (diagonal straight line) and $e = 0.7$ (curved line). The
  marked points are defined in the text.}
\label{fig60}
\end{figure}

Use the notation $E = f_\MtoE(M,e)$ to describe the transformation
between $M$ and $E$ and similar notation for the transformations
between other pairs of angles.  The main transformations of interest
are $f_\MtoT$ and $f_\TtoM$.  The function $f_\MtoT$ here plays the
role of the function $h$ in Section \ref{sec:back:ekfs} and takes an angle
on the mean anomaly scale to an angle on the true anomaly scale.
Fig. \ref{fig60} shows a plot of $T$ vs. $M$ for $e = 0.7$.  Notice
the highly nonlinear behaviour.

The mean anomaly is the simplest angle mathematically and statistically
because it changes at a constant rate in time:
$$
\phi(t) = \phi(0) + t n,
$$
where $n$ is the mean motion.  The actual angular position along the
orbit is given by the true anomaly $\theta(t) = f_\MtoT(\phi(t))$. 

For the purposes of this section, suppose that the initial mean
anomaly $\phi(0)$ at time $t=0$ is known exactly, but that the mean
motion $n$ has some Gaussian uncertainty, $ n \sim N(\mu_n, \sigma^2_n)$.
Then after some time $t_1$, say, the mean anomaly has distribution
$$
\phi(t_1) \sim N(\phi(0)+t_1\mu_n , \,t_1 \sigma_n^2).
$$
However, the observation is on the true anomaly scale
$$
\theta_\obs \sim N(\theta(t_1),\tau^2), \quad \theta(t_1) = f_\MtoT(\phi(t_1)).
$$

Note that even if $\sigma^2_n$ is small, $\sigma^2= t_1 \sigma_n^2$
can still become large by considering a large propagation time $t_1$.
For the purposes of this paper suppose $\sigma^2$ is not too large
in order to avoid winding number issues.  In particular,  restrict 
$\sigma = t_1^{1/2} \sigma_n$ to be  substantially less than 360$^o$ so
that $\theta_\obs$ can be treated as a number unambiguously satisfying
$|\theta_\obs - \theta(t_1)| < 360^o$.  In other words the number of
whole orbits undergone is essentially known.  The choices
$\sigma = 25^o$ and $\sigma = 15^o$ are used in the examples below.

At the same time, the typical angles-only observations will be highly
accurate.  Three choices for $\tau$ are used in each example: (a)
$\tau=0^o$ for a perfect measurement, (b) $\tau=5.5E-04^o$ (equal to 2
arc-seconds) for a realistic measurement error, and (c) $\tau=2^o$ for a
good but less accurate measurement.

\nil{
The larger value is far larger than typically encountered
in practice, yet the discrepancies between the various filters still
stand out prominently.
}

For convenience suppose time is measured in orbits, so that $t=1$
denotes the time needed for one complete orbit. Also, to help our
intuition, measure angles in degrees.  Finally note that estimating
$n$ is equivalent to estimating $\phi(t_1)$ in this setting.  Let
$\phi(t_1)$ here correspond to the state $x$ in Section
\ref{sec:back:classic}, with variance $\sigma^2$, and let $z_\obs$
denote the observed true anomaly, with variance $\tau^2$.  Thus the
the state variable is the mean anomaly $x = \phi(t_1)$ lying on the
signal scale, and the observation is the true anomaly
$z_\obs = \theta_\obs$ lying on the measurement scale.

Here are two numerical examples to illustrate the pitfalls of the EKF,
UKF and to demonstrate the benefits of the IEKF, IUKF, OCEKF and
OCUKF. For both examples a high value of ellipticity is used, $e=0.7$,
so that the function $h = f_\MtoT$ is very nonlinear and the
differences between the various filters stand out prominently.  The
parameters for each example are listed in Table
\ref{table:table:parameters} and highlighted in Figure \ref{fig60}.
In this figure, the projection of the circles onto the horizontal axis
gives the prior means $\mu_x$ for Examples 1 and 2, respectively; the
projection of the squares onto the vertical axis gives the corresponding
observations $z_\obs$, and the projection of the squares onto the
horizontal axis gives the corresponding values of
$x_\obs = h^{-1}(z_\obs)$.

For each example three choices are considered for the measurement
standard error: (a) zero, $\tau=0^o$, (b) ``small'', $\tau=5.5E-04^o$
= 2 arc seconds, and (c) ``large'', $\tau = 2^o$.  The posterior means
and standard deviations for various filters are summarized in Table
\ref{table:ex12}.  The row labelled ``Exact'' in that table gives the
exact moments from the true posterior distribution, as computed by
numerical integration.

\emph{Example 1.} Since 
$x_\obs = 310^o = 260^o + 2 \times 25^o = \mu_x + 2 \sigma$,  the
observation is mildly unusual but not infeasible under the prior
distribution.  However, the different filters produce wildly different
results.

\begin{figure}[t!]
\includegraphics[width=8.2cm, keepaspectratio]{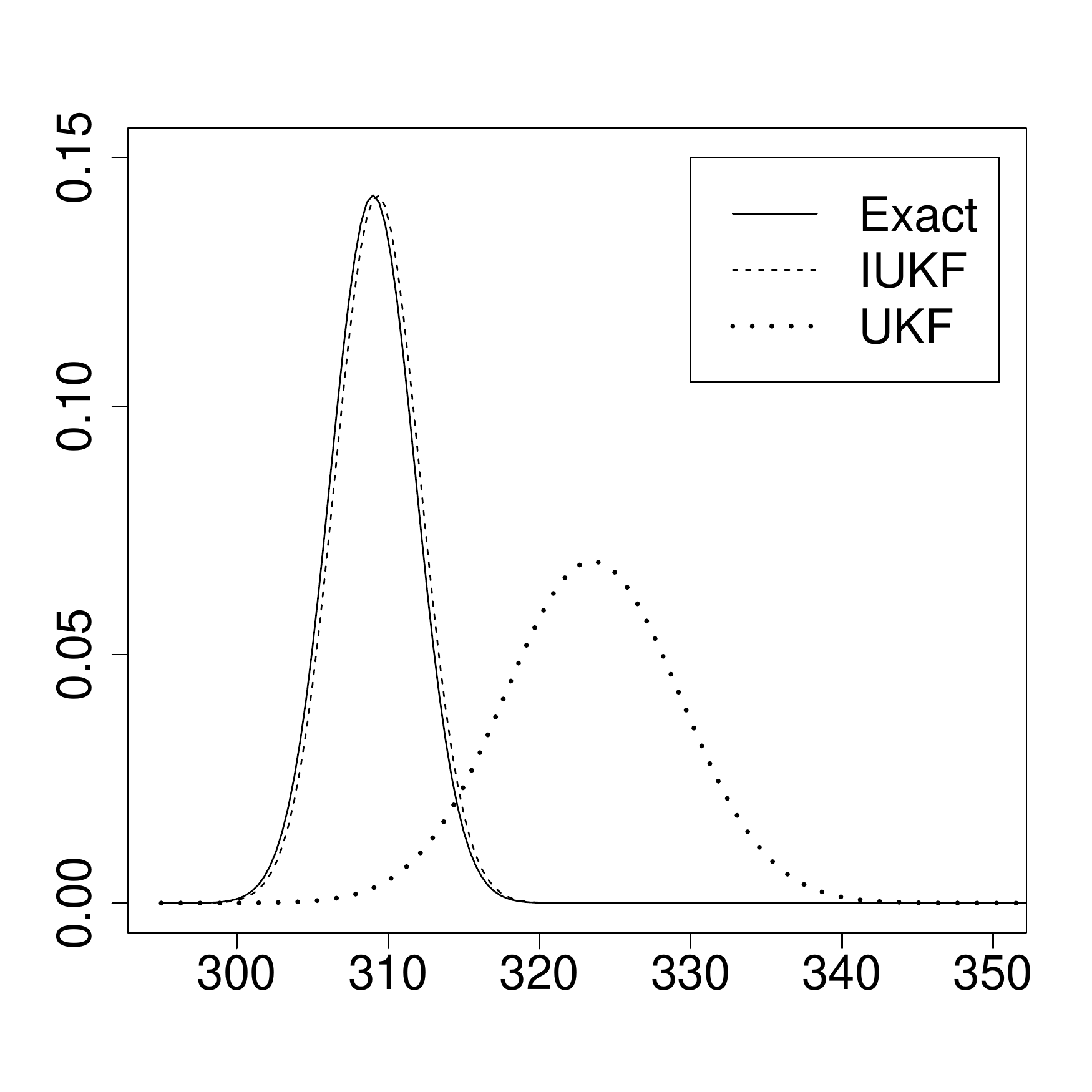}
\caption{Example 1(c). Exact posterior density together with the
  approximating Gaussian density from the IUKF and UKF filters. The
  exact posterior and the IUKF posterior densities are virtually
  indistinguishable.}
\label{fig61}
\end{figure}

For all three choices of $\tau$, the posterior means from the
IEKF,IUKF, OCEKF and OCUKF filters are either identical or very close
to each other and to the exact value.  However, the EKF and UKF
filters produce posterior means that are so different, they are
essentially incompatible with the exact posterior distribution,
The discrepency becomes greater as $\tau$ gets smaller.
Fig. \ref{fig61}) illustrates the issue in the least
discrepant situation, case (c) with $\tau=2^o$.  The
UKF posterior mean $323.5^o$ is more than five standard deviations
from the exact posterior mean $309.0^o$ under the exact posterior
distribution, $(323.5^o-309.0^o)/2.8^o = 5.18$.  The EKF is even worse.

Similarly, with a small exception, the posterior standard deviations
from the IEKF,IUKF, OCEKF and OCUKF filters are either identical or
very close to each other and to the exact value. The exception is the
IUKF and OCUKF for $\tau=0^o$.  In this case the true posterior is a
point mass at the posterior mean with posterior standard deviation
$0^o$.  However, the posterior standard deviation is overestimated by
the IUKF and OCUKF filters.  The reason is that these filters use
differences rather than derivatives to deal with the nonlinearity.
The problem would be ameliorated by using a smaller tuning parameter $\alpha$
in the construction of the unscented filters.

\emph{Example 2.} As in Example 1,
$x_\obs = 65^o = 35^o + 2 \times 15^o = \mu_x + 2 \sigma$, so again
the observation is mildly unusual but not infeasible under the prior
distribution. For the most part, the comparison between the different
filters is very similar to the comparison in Example 1.  The main
difference is in the direction of the error.  In Example 1 the
posterior means for the EKF and UKF are \emph{larger} than the true
posterior mean, whereas in Example 2, the posterior means for the EKF
and UKF are \emph{smaller} than the true posterior mean.  The reason
is that $h$ in Fig. \ref{fig60} is \emph{convex} at $\mu_x$ in Example
1 and it is \emph{concave} at $\mu_x$ in Example 2.

\begin{table}[ht!]
\begin{center}
\caption{The prior mean $\mu_x$ and its standard deviation $\sigma$, plus the observation $z_{obs}$ and its standard 
deviation $\tau$, for Examples 1 and 2.}
\label{table:table:parameters}
\begin{tabular}{cc}
\hline
Example 1 & Example 2 \\ \hline
$\mu_x = 260^{o}$ & $\mu_x = 35^{o}$ \\
$\sigma = 25^{o}$ & $\sigma = 15^{o}$ \\
$z_{obs} = 225.5^{o}$ & $z_{obs} = 143.6^{o}$\\
$x_\obs = 310^o$ & $x_\obs = 65^o$\\
\multicolumn{2}{c}{a: $\tau=0^o$} \\
\multicolumn{2}{c}{   b: $\tau=5.5E-04^o$} \\
\multicolumn{2}{c}{c: $\tau=2^o$} \\ \hline
\end{tabular} \end{center} 
\end{table}

\begin{table}[ht!]
\begin{center}
\caption{Posterior means and standard deviations from various
  filters for Examples 1(a, b, c) and 2(a, b, c).}
\label{table:ex12}
\begin{tabular}{cccccccc}
\hline
KF/Example & moment & 1(a) & 1(b) & 1(c) & 2(a) & 2(b) & 2(c) \\ \hline 
Exact  & mean  & 310.0$^{o}$     & 309.9999$^{o}$    & 309.0$^{o}$      & 65.0$^{o}$       & 64.9702$^{o}$ 	& 63.5$^{o}$ \\ 
& s.d.  & 0$^{o}$ 	&     7.7E-04$^{o}$  	& 2.8$^{o}$   	& 0$^{o}$ 	& 1.1E-03$^{o}$ 	&  3.5$^{o}$ \\ 
EKF  & mean  & 329.8$^{o}$     & 329.8584$^{o}$    & 326.1$^{o}$      & 55.0$^{o}$       & 55.0748$^{o}$ 	& 54.8$^{o}$ \\ 
& s.d.  & 0$^{o}$ 	& 1.6E-03$^{o}$  	& 5.7$^{o}$   	& 0$^{o}$ 	& 4.9E-04$^{o}$ 	&  1.7$^{o}$ \\ 
UKF  & mean  & 327.1$^{o}$     & 325.2281$^{o}$   & 323.5$^{o}$      & 59.1$^{o}$       & 59.1864$^{o}$ 	& 58.9$^{o}$ \\ 
& s.d.  & 2E-03$^{o}$ 	& 1.6E-03$^{o}$ 	& 5.8$^{o}$  	& 4E-03$^{o}$ 	& 5E-04$^{o}$ 		& 1.8$^{o}$  \\
IEKF & mean  & 310.0$^{o}$     & 309.9999$^{o}$   & 309.3$^{o}$      & 65$^{o}$       & 64.9702$^{o}$ 	& 63.2$^{o}$ \\
& s.d.  & 0$^{o}$ 	& 7.7E-04$^{o}$ 	& 2.8$^{o}$  	& 0$^{o}$ 	& 1.1E-03$^{o}$ 	& 3.5$^{o}$  \\
IUKF & mean  & 310.0$^{o}$     & 309.9999$^{o}$   & 309.3$^{o}$      & 65$^{o}$       & 64.9702$^{o}$ 	& 63.2$^{o}$ \\
& s.d.  & 7E-03$^{o}$ 	& 7.7E-04$^{o}$ 	& 2.8$^{o}$  	& 2E-03$^{o}$ 	& 1.1E-03$^{o}$ 	& 3.5$^{o}$  \\
OCEKF & mean  & 310.0$^{o}$     & 309.9999$^{o}$   & 309.3$^{o}$      & 65$^{o}$       & 64.9702$^{o}$ 	& 63.1$^{o}$ \\
  & s.d.  & 0$^{o}$ 	& 7.7E-04$^{o}$ 	& 2.7$^{o}$  	& 0$^{o}$ 	& 1.1E-03$^{o}$ 	&  3.7$^{o}$ \\ 
OCUKF & mean  & 310.0$^{o}$     & 309.9999$^{o}$   & 309.3$^{o}$      & 65$^{o}$       & 64.9702$^{o}$ 	& 63.2$^{o}$ \\
  & s.d.  & 7E-03$^{o}$ & 7.7E-04$^{o}$ 	& 2.8$^{o}$  	& 2E-03$^{o}$ 	& 1.2E-03$^{o}$ 	&  3.7$^{o}$ \\ \hline
\end{tabular} \end{center} 
\end{table}

\section{Conclusions}
\label{sec:concl}

Several conclusions can be made from these and other simulations.
\begin{itemize}
\item The EKF and UKF are standard methods to deal with nonlinear
  filtering problems; the UKF in particular is often used as an
  ``off-the-shelf'' method.  However, when the nonlinearity is high
  and the observation variance $\tau^2$ is small relative to the prior
  state variance $\sigma^2$, these filters can perform very poorly.

\item The development of the OCEKF and OCUKF in Sections
  \ref{sec:back} and \ref{sec:intuition} makes it clear why this
  behavior occurs.  When $\tau^2$ is small, it is better to base the
  Taylor series expansion at or near the observation, rather than at
  the prior mean.

\item In terms of performance, the IEKF and OCEKF are
  very similar for the examples in this paper.  Further, the posterior
  means and variances computed using these filters closely match the
  true posterior moments. 

\item The same statement is true for the IUKF and OCUKF, with one
  proviso.  If $\tau^2$ is very small or 0, the value of the posterior
  variance will depend noticeably on the tuning parameter $\alpha$.
  To ensure the posterior variance from the filter is close to the
  true posterior variance, it is necessary to reduce the value of
  $\alpha$ below the default choice of $\alpha = 0.001$.  Indeed, as
  pointed out in Section \ref{sec:back:ukfs}, the limiting values of
  the IUKF and OCUKF as $\alpha \rightarrow 0$ are just the IEKF and
  OCEKF.

\item One advantage of the OCEKF and OCUKF over the IEKF and the IUKF
  is that they do not require iteration.

\item However, an advantage of the IEKF and the IUKF over the OCEKF
  and the OCUKF is that they are more widely applicable.  In
  situations where $\tau^2$ is not small relative to $\sigma^2$
  (simulations not shown here), the posterior moments from the IEKF
  and IUKF can be closer to the true posterior moments.

\item In the literature several other versions of the UKF and IUKF
  have been defined (e.g.~\cite{CDKF1, all-ukf, IUKF_bad1,
    IUKF_bad2, all-bad4, all-bad5, all-bad6, all-bad7, all-bad8}).  However, in the setting of this paper, they are all
  identical to either the UKF or the IUKF.

\end{itemize}

\section*{Acknowledgment}
This material is based upon work supported by the Air Force Office of
Scientific Research, Air Force Materiel Command, USAF under Award
No. FA9550-19-1-7000.

\end{document}